\documentclass[%
reprint,
superscriptaddress,
amsmath,amssymb,
aps,
pra,
]{revtex4-2}

\usepackage{graphicx}
\graphicspath{ {Figures/} }

\usepackage{dcolumn}
\usepackage{bm}
\usepackage[export]{adjustbox}
\usepackage[normalem]{ulem}
\usepackage{comment}

\usepackage{mathtools} 
\DeclarePairedDelimiter\bra{\langle}{\rvert}
\DeclarePairedDelimiter\ket{\lvert}{\rangle}
\DeclarePairedDelimiterX\braket[2]{\langle}{\rangle}{#1 \delimsize\vert #2}

\usepackage[utf8]{inputenc}
\usepackage[T1]{fontenc}

\usepackage{mathrsfs}
\usepackage{dsfont}

\usepackage{amssymb}
\usepackage{amsmath}
\usepackage{amsfonts}
\usepackage{amsthm}
\usepackage{float}

\newcounter{rem}
\newcommand{\mc}[1]{\mathcal{#1}}

\def\>{\rangle}
\def\<{\langle}
\newcommand{\proj}[1]{| #1 \rangle\! \langle #1 |}
\newcommand{\idty}{\mathds{1}}
\def\tr{{\rm tr}}
\def\pr{{\rm Pr}}

\def\rho{{\varrho}}

\def\textbf#1{{\bf #1}}
\newcommand{\Cx}{\mathbb{C}}

\newcommand{\Rl}{\mathds{R}}
\gdef\lvert{\delimiter"426A30C }
\gdef\rvert{\delimiter"526A30C }

\usepackage{xcolor}

\usepackage{hyperref}
\hypersetup{pdfpagemode=UseNone}
\usepackage{cleveref}

\begin{document}
	
\title{Boltzmann counting in  Hilbert space}

\author{Ra\'ul O. Vallejos}
\email{vallejos@cbpf.br}
\affiliation{Centro Brasileiro de Pesquisas F\'{\i}sicas, Rua Dr. Xavier Sigaud, 150, Rio de Janeiro, RJ, Brazil}
\author{Isadora Veeren}
\email{isadora.veeren@polytechnique.edu}
\affiliation{CPHT, LIX, CNRS, Inria, École polytechnique, Institut Polytechnique de Paris, Palaiseau, France}
\author{Frederico Brito}
\email{frederico.brito@tii.ae}
\affiliation{Quantum Research Center, Technology Innovation Institute, Abu Dhabi, UAE}
\author{Fernando de Melo}
\email{fmelo@cbpf.br}
\affiliation{Centro Brasileiro de Pesquisas F\'{\i}sicas, Rua Dr. Xavier Sigaud, 150, Rio de Janeiro, RJ, Brazil}

\date{\today}	
 

\begin{abstract}
We introduce a geometric entropy for quantum preparations, defined as the logarithm of the Hilbert-space volume of pure states compatible with a given set of constraints. This construction extends Boltzmann’s counting perspective to the quantum setting, where compatible states need not be orthogonal and the relevant notion of “number of states” is naturally replaced by a volume in state space. We analyze three classes of constraints: restriction to a subspace, fixed expectation values, and coarse-grained subsystem descriptions. For representative examples, including subspace projection, spin expectation values, partial trace, and an imperfect detector map, we obtain explicit scaling laws and closed-form expressions for the associated volumes. The resulting framework provides a geometric measure of quantum ignorance at the level of the preparation and complements entropy notions based on density matrices and coarse graining.
\end{abstract}

\maketitle

\section{Introduction}

To prepare a physical system means to measure it. To prepare a gas in one half of a box, one has, for instance, to load the gas in the box and push a piston up to the middle of the box. In doing so, we gain information, that is, we perform a measurement on the (coarse-grained) position of the atoms.

Of course, this preparation of a macroscopic physical property does not tell us the individual positions of the atoms or their velocities. So, given a macroscopic preparation of a physical system, what microscopic description should we assign to it? And how should we quantify our ignorance about the microscopic state? These are foundational questions of statistical physics, both in its classical and quantum versions.

Classically, the whole scene unfolds in phase space, with a macroscopic preparation naturally defining a set of compatible microstates. This set plays a dual role: on the one hand, its volume determines the Boltzmann entropy, defined as the logarithm of the phase-space measure of the corresponding macroregion. On the other hand, it provides the basis for constructing statistical ensembles, most notably through the assignment of probability distributions over the compatible microstates.

In the quantum domain, however, this picture becomes less direct. A preparation is again associated with a restriction on the possible microscopic states. However, due to the uncertainty principle, points in phase space are no longer physically meaningful, and a direct phase-space description is no longer immediate. The natural arena is instead Hilbert space, whose points represent pure states associated with microscopic configurations. Moreover, while statistical states are naturally described by density matrices and entropy is quantified by the von Neumann entropy, the relation between these constructions and an underlying set of compatible pure states is less transparent.

This raises the question of whether the classical structure — where both statistical states and entropy originate from a set of compatible microstates — can be meaningfully extended to the quantum setting. In particular, can one define both an assignment of statistical states and a notion of entropy directly from the set of pure states compatible with a given preparation?

In previous work \cite{correia2021,vallejos2022}, we addressed the first of these questions by introducing an average assignment map, which associates to a given preparation the convex combination of all compatible pure states. This construction provides a direct assignment of a statistical state based solely on the underlying set of microstates, without invoking an optimization principle such as maximum entropy. Interestingly, this assignment does not, in general, coincide with the one obtained via the maximum entropy principle. In the present work, we complement this construction by proposing a corresponding notion of entropy, defined as the logarithm of the volume of the set of compatible pure states. This definition can be seen as a quantum analogue of Boltzmann's entropy, formulated directly at the level of the preparation. In this sense, the entropy introduced here quantifies the intrinsic size of the quantum state space selected by a preparation constraint.

A related geometric approach to quantum ignorance has recently been proposed in Ref.~\cite{Ray2023}, which assigns a volume to the set of purifications compatible with a reduced state. Our construction is similar in spirit, but it assigns volume directly to the set of pure states compatible with a preparation.

In the past several years, a related quantity, known as the observational entropy, has been developed \cite{Safranek2020,Safranek2021,Safranek2019a,Safranek2019b} into a broad framework for coarse-grained notions of quantum and classical entropies. It is defined relative to a chosen coarse-graining and therefore captures the uncertainty associated with limited observational access. The entropy introduced in this work is conceptually different: it is defined directly from the set of pure states compatible with a given preparation, and its value is the logarithm of the corresponding Hilbert-space volume. Thus, it may be viewed as a Boltzmann-type entropy for quantum preparations, complementing observational entropy in the broader landscape of quantum coarse-grained descriptions.

\section{Quantum preparations: set of compatible states}

We start by defining the set of compatible microscopic states for different types of preparations.

\subsection{Subspace restriction}

One of the most prevalent preparations of physical systems is by restricting the possible values of a given quantity. For instance, one can say that the energy of the system has a fixed value, which then leads to the microcanonical ensemble. For a system with associated Hilbert space $\mc{H}_D$, with $\dim(\mc{H}_D)=D$, a restriction defines a subspace $\mc{H}_R\subseteq \mc{H}$, with $\dim(\mc{H}_R)=D_R\le D$ \cite{Popescu2006}.

Let $\Pi_R$ be the projector onto $\mc{H}_R$. The set of microscopic states compatible with this preparation is then:
\begin{equation}
\Omega_R=\{ \ket{\psi}\in \mc{H}_D \; |\; \Pi_R \ket{\psi} = \ket{\psi}\}.
\end{equation}

This is a vector restriction, which implies a series of equalities. Let $\Pi_R^\perp= \idty - \Pi_R$ be the orthogonal projector to the subspace $\mc{H}_R$, then the subspace constraint reads $\Pi_R^\perp \ket{\psi}=0$. Writing the full space as $\mc{H}=\mc{H}_R\oplus \mc{H}_R^\perp$, and letting $\{\ket{e_i}\}_{i=1}^{D_R}$ and $\{\ket{e_i^\perp}\}_{i=D_R+1}^{D}$ be orthonormal bases for the two subspaces, then 
\begin{equation}
\ket{\psi}=\sum_{i=1}^{D_R} c_i \ket{e_i}+\sum_{j=D_R+1}^{D} c_j^\perp \ket{e_j^\perp},
\end{equation}
with $c_i,c_j^\perp \in \Cx$ and $\sum_{i=1}^{D_r}|c_i|^2+\sum_{j=D_R+1}^{D} |c_j^\perp|^2=1$. With all that, the set of compatible states with a projection can be defined as:
\begin{equation}
\Omega_R=\left\{ \ket{\psi}\in \mc{H}_D \; |\; c_i^\perp =0, \forall i\in\{D_R+1,\ldots, D\}\right\}.
\end{equation}

It should be realized that as the $c_i^\perp$ are complex numbers, then these are $2(D-D_R)$ equality constraints.

\subsection{Fixed expectation value}
\label{sec:fix-exp-value}

Another usual type of restriction is by fixing the expectation value of observable quantities. A paradigmatic example here is the construction of the canonical ensemble, where only the expectation value of the system's energy is fixed.

In general, for a given observable $A$ acting on $\mc{H}_D$, if we fix its expectation value to $a\in \Rl$, then the set of compatible states is described as:
\begin{equation}
\Omega_a = \left\{ \ket{\psi}\in \mc{H}_D \; |\; \<\psi|A|\psi\> = a\right \}.
\label{eq:Omega-a-00}
\end{equation}

Notice that this set does not form a subspace. It is a quadratic manifold on the space of states.

If one simultaneously fixes the expectation value of $N$ observables $A_i$ to the respective values $a_i$, then the set of compatible underlying states is 
\begin{equation}
\Omega_{\vec{a}} = \left\{ \ket{\psi}\in \mc{H}_D \; |\; \<\psi|A_i|\psi\> = a_i,\; \forall i\in [N] \right\}.
\label{eq:Omega-a-vec}
\end{equation}

If the expectation value of more observables is fixed, the order in which this is done does not matter  and no commutation relation among the observables is assumed. Nevertheless, each new constraint might reduce the set dimension by one, and the set might eventually become empty.

Several works have investigated ensembles of pure states constrained by a fixed expectation value. In particular, Müller, Gross, and Eisert analyzed the geometry of this manifold and established concentration-of-measure results for observables restricted to it \cite{MullerGrossEisert2011}. Related dynamical questions were considered by Bartsch and Gemmer, who showed that states drawn from expectation-value constrained ensembles typically exhibit very similar time evolution \cite{BartschGemmer2009}. Reimann later generalized and unified these ideas, clarifying the conditions under which dynamical typicality emerges for isolated many-body systems and relating the constrained ensemble to its thermodynamic interpretation \cite{Reimann2018}.

\subsection{Fixed subsystem}

Another possible preparation is when one only has access to some degrees of freedom of the total system, which can be understood as a generalized subsystem. One common example of this type of preparation is the traditional situation in open quantum systems, where one has control over a small subsystem that interacts with, in principle, an uncontrollable environment.

In general, subsystems are defined by the action of coarse-graining channels, which are dimension-reducing completely positive and trace-preserving (CPTP) channels. Let $\Lambda: \mc{H}_D \mapsto \mc{H}_d$, with $\dim(\mc{H}_d)=d<D$, be a CPTP channel, that is, let $\Lambda$ be a coarse-graining map such that the underlying total system has dimension $D$ and the effective subsystem has dimension $d$. If one is able to prepare the state of the subsystem to be $\rho$, then the compatible microscopic set is given by
\begin{equation}
\Omega_{\Lambda}(\rho) = \left\{ \ket{\psi}\in \mc{H}_D \; \big| \; \Lambda(\proj{\psi})= \rho \right\}.
\end{equation}

Notice that this type of preparation is, in fact, a particular case of the previous one. Indeed, let $\{\sigma_i\}_{i=1}^{d^2-1}$ be Hermitian observables forming an orthonormal basis of the space of traceless operators on $\mathcal H_d$, normalized as
\begin{equation}
\operatorname{tr}(\sigma_i\sigma_j)=d\,\delta_{ij}.
\end{equation}
Together with the identity operator, the set $\{\sigma_i\}_{i=1}^{d^2-1}$ forms a basis of Hermitian operators on $\mathcal H_d$. Then the condition $\Lambda(\ket{\psi}\bra{\psi})=\rho$ can be equivalently expressed as the set of $d^2-1$ expectation-value constraints
\begin{equation}
\langle\psi|\Lambda^*(\sigma_i)|\psi\rangle=\operatorname{tr}(\sigma_i\rho), \qquad i=1,\dots,d^2-1.
\end{equation}
where $\Lambda^*$ is the dual map of $\Lambda$. The identity constraint is automatically satisfied because both sides have unit trace.

However, the converse does not hold in general: not every collection of $d^2-1$ fixed expectation values corresponds to the reduced description of a $d$-dimensional subsystem. Such a description must arise from a physical coarse-graining map, so the constraints must satisfy additional consistency conditions.
\\
\\
Of course there can be other types of preparations, for instance combinations of the above-mentioned procedures.

\section{Volume of microscopic compatible states}

Now that the sets of microscopic states compatible with a given preparation are defined, we can count how many states are in such sets. That is, we can determine their volumes.

Like in classical statistical mechanics, in general, associated to a given preparation there will be infinitely many compatible microscopic states. Instead of coarse-graining the space of states, as is routinely done in standard classical statistical mechanics, here we take the more modern and rigorous measure-theoretic approach to weigh such sets. The idea is to evaluate the fraction of states that would belong to a set $\Omega$ if pure states from $\mc{H}_D$ are sampled uniformly from the space of states, that is, according to the Haar measure.

Nevertheless, as the compatible sets $\Omega$ are defined by imposing equality constraints over the full set of states, the probability of obtaining a state that exactly fulfills the constraints is zero when sampling from the Haar measure on $\mc{H}_D$
. In other words, these sets have zero volume.

\begin{widetext}

To deal with such an issue, we need to define regularized versions of the compatible sets. For the main types of compatible sets discussed above, their regularized versions are
\begin{equation}
\begin{split}
\Omega_R^\epsilon &= \left\{ \ket{\psi}\in \mc{H}_D \; \middle|\; |\Re(c_i^\perp)| \le \epsilon \text{ and } |\Im(c_i^\perp)| \le \epsilon,\ \forall i\in\{D_R+1,\ldots, D\}\right\}, \\
\Omega_{\vec{a}}^\epsilon &= \left\{ \ket{\psi}\in \mc{H}_D \; \middle|\; |\langle\psi|A_i|\psi\rangle - a_i| \le \epsilon,\ \forall i\in[N] \right\}, \\
\Omega_{\Lambda}(\rho) &= \left\{ \ket{\psi}\in \mc{H}_D \; \middle|\; |\langle\psi|\Lambda^*(\sigma_i)|\psi\rangle - \tr(\sigma_i \rho)| \le \epsilon,\ \forall i\in [d^2-1] \right\}.
\end{split}
\label{eq:regularized-sets}
\end{equation}
where $\epsilon>0$ controls the tolerated error on the constraint.

\end{widetext}

In general, for an $\epsilon$-regularized compatible set defined as
\begin{equation}
\Omega_{\vec{f}}^\epsilon = \left\{ \ket{\psi} \in \mc{H}_D \; \middle|\; |f_i(\psi)| \le \epsilon,\ \forall i\in [K]\right\},
\label{eq:omegaf-eps}
\end{equation}
with $f_i:\mc{H}_D \mapsto \Rl$, the volume $|\Omega_{\vec{f}}^\epsilon|$ of such a set can be evaluated as
\begin{equation}
|\Omega_{\vec{f}}^\epsilon|
= \int_{\Omega_{\vec{f}}^\epsilon} d\psi
= \int d\psi \;\prod_{i=1}^{K}\Theta\!\left(\epsilon-|f_i(\psi)|\right),
\label{eq:volume_def}
\end{equation}
where $\Theta$ is the Heaviside step function, and both integrals are performed over the Haar measure. This is the main object of interest in our work.

As the Heaviside step function can be understood as an indicator function, the volume definition is equivalent to the probability that a sampled pure state $\ket{\psi}$ satisfies all the constraints simultaneously. In mathematical terms,
\begin{equation}
|\Omega_{\vec{f}}^\epsilon|
= \pr_{\ket{\psi}\sim \text{Haar}}\Big(|f_i(\psi)| \le \epsilon,\ \forall i\in[K]\Big).
\label{eq:prob-volume}
\end{equation}
In this way, the volume defined here varies from 0 to 1, with its maximum value attained when no restriction is imposed on the set of states.

For $\epsilon \ll 1$, one may use the standard approximation
\begin{equation}
\Theta(\epsilon-|x|) \sim 2\epsilon\,\delta(x),
\label{eq:theta-delta}
\end{equation}
which corresponds to the familiar rectangular approximation of the Dirac delta function.

The general expression for the volume can then be written as
\begin{equation}
|\Omega_{\vec{f}}^\epsilon|
\approx (2\epsilon)^K \int d\psi \prod_{i=1}^K \delta(f_i(\psi)).
\label{eq:omega-epsilon}
\end{equation}
As such, the product of delta functions can be seen as the density of states satisfying the constraints, or, geometrically, as the measure of the compatible surface, while the term proportional to $\epsilon^K$ gives the thickness of the compatible shell.

Volume regularization is also important in numerical computations, for example in rejection-sampling schemes. One may wish to check an analytical result numerically, or the analytical calculation may be too difficult to carry out in closed form. In such cases, a numerical method may be the only practical alternative. For these methods to be well defined, the constraint manifold must have finite volume, which in practice means endowing it with a small but nonzero thickness.

To finish this general section, we can now define a quantity analogous to Boltzmann's entropy, which we call Boltzmann's quantum entropy:
\begin{equation}
S_Q(\Omega_{\vec{f}}^\epsilon)
= \log\!\big(V_0\,|\Omega_{\vec{f}}^\epsilon|\big),
\label{eq:q_boltzmann}
\end{equation}
where $V_0$ is an arbitrary number larger than one that sets the scale for the unconstrained volume. In other words, when no constraint is imposed, the entropy is $\log(V_0)$, its maximum value. When constraints are imposed on the system's preparation, the quantity $|\Omega_{\vec{f}}^\epsilon|$ becomes smaller than one, meaning that we have more information about the microscopic state of the system, and thus $S_Q$ decreases. The specific value of $V_0$ is not important, as one is usually interested in entropy differences between two situations. For the same reason, when the number of constraints is fixed, the precise size of $\epsilon$ is not important, and only the density of states matters.

We note that a related geometric perspective on quantum ignorance was proposed in Ref.~\cite{Ray2023}, where the authors define a metric-induced volume on the manifold of purifications of a reduced density operator and interpret it as a measure of missing information. While that construction is based on purification manifolds associated with reduced states, the present approach assigns volume directly to the set of pure states compatible with a given preparation. The two frameworks are therefore similar in their geometric motivation, but differ in both the notion of compatibility and the measure used to quantify it.

In what follows, we evaluate the volume of compatible states for several representative preparations. More precisely, we compute the measure of the corresponding constraint manifolds, i.e., the volume density. For brevity, however, we refer to these quantities as volumes, denoted by $\mc{V}$, with the understanding that the regularized volume is obtained by multiplying the manifold measure by a thickness of order $\epsilon^K$.

\subsection{Fixed expectation values}

\subsubsection{Single fixed expectation value}

Here we discuss the volume
\begin{equation}
    \mathcal{V}_a = \int d\psi \,\delta(\langle \psi|A|\psi\rangle - a),
\label{eq:volume-a}
\end{equation}
where the integral is over pure states in $\mc{H}_D$ with respect to the Haar measure, $A$ is a Hermitian operator, and $a$ is a real number.

This quantity was studied by Brody et al.~\cite{Brody1998,Bender2005,Brody2007,Brody2007b}, who evaluated $\mathcal{V}_E$ in the special case where $A$ is the Hamiltonian and $a=E$, in connection with a proposed redefinition of the microcanonical ensemble. In that framework, the set of compatible states is defined by fixing the energy expectation value rather than by restricting to an energy eigenspace.

The same integral may also be interpreted as the probability density of the random variable $a=\langle \psi|A|\psi\rangle$ when $|\psi\rangle$ is drawn uniformly from the pure-state manifold. This point of view has been developed in detail in the mathematical physics literature, where the dependence of the resulting density on both the spectrum of $A$ and the value of $a$ has been extensively 
analyzed~\cite{Dunkl2011,Dunkl2011b,Puchal2012}.

Venuti and Zanardi~\cite{Venuti2013} also consider ensembles defined by fixed expectation values, 
derive the corresponding probability distributions, and discuss the statistical mechanics associated 
with the generalized microcanonical ensemble of Brody et al. 
In particular, they obtain the exact Haar-induced probability density \(\mc{V}_A(a)\)  for the expectation value of a Hermitian observable over random pure states, showing that it is supported on the line segment spanned by the eigenvalues of \(A\), and it is generally a piecewise polynomial determined by the eigenvalues 
and their degeneracies (see also \cite{Dunkl2011}). 
They provide closed expressions for both non-degenerate and degenerate spectra, recover beta distributions 
in special cases such as projectors (see Sec.~\ref{sec:subs-proy}), and show that in large Hilbert-space dimension the distribution becomes sharply concentrated with an asymptotic Gaussian limit for suitably centered and rescaled observables. 
These results give an exact finite-dimensional description of typical expectation values and connect naturally to concentration-of-measure and central-limit behavior in high-dimensional quantum systems.

\subsubsection{Various fixed expectation values}

To the best of our knowledge, there are no general results concerning the volume of the sets
$\Omega_{\vec{a}}$, defined in Eq.~\eqref{eq:Omega-a-vec}.
Nevertheless, two of the present authors studied particular instances in \cite{correia2021,vallejos2022}
in the context of coarse-graining maps (Sec.~\ref{sec:coarse-graining-maps}).
As an illustrative example, consider a family of compatible sets for which the dimension $D$ of the Hilbert space can be made arbitrarily large:
\begin{equation}
\Omega_{\vec{J}} =
\left\{ \ket{\psi}\in \mc{H}_D \;\middle|\;
\left\langle \psi \left| \frac{J_i}{j} \right| \psi \right\rangle = r_i,\quad i\in[3]
\right\},
\label{eq:Omega-J-vec}
\end{equation}
where $\vec{J}=(J_1,J_2,J_3)$ denotes the angular-momentum operator on an irreducible spin-$j$ block, with block dimension $D=2j+1$, and $\vec{r}$ is a vector with $|\vec{r}|\leq 1$, which may be regarded as a Bloch vector.
The coarse-graining channel associated with the set $\Omega_{\vec{J}}$ was first introduced in Ref.~\cite{ibrahim}.

\begin{figure}[t]
    \centering
    \includegraphics[width=\columnwidth]{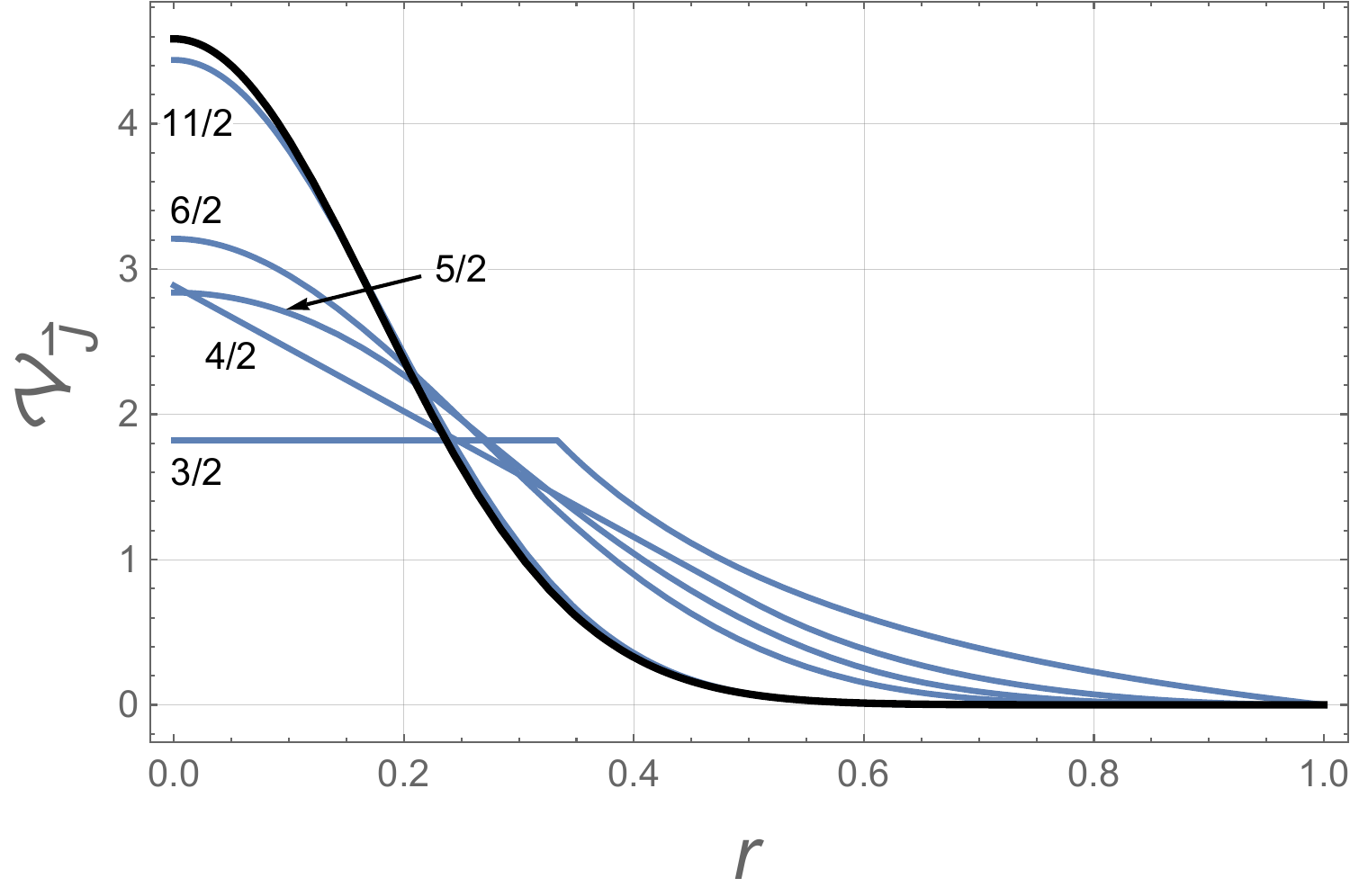}
    \caption{
Comparison between the exact distributions $\mathcal{V}_{\vec{J}}$ (blue) and the Gaussian approximation (black).
Exact curves correspond to $j=3/2,4/2,5/2,6/2,11/2$, with dimensions $D=4,5,6,7,12$, respectively.
The Gaussian fit uses the unidimensional normalization of the radial profile, so that its width parameter satisfies $\sigma^2=1/[3(D-1)]$.
Gaussianity of the exact curves grows with $D$. 
}
    \label{fig:volume-j}
\end{figure}

The volume $\mathcal{V}_{\vec{J}}$ of the set $\Omega_{\vec{J}}$ was calculated in Ref.~\cite{vallejos2022}. By rotational symmetry, $\mathcal{V}_{\vec{J}}$ depends only on $r=|\vec{r}|$. Moreover, it is a piecewise polynomial, with both the number of pieces and the polynomial degree increasing with the dimension $D$.
For moderately large $D$, the function $\mathcal{V}_{\vec{J}}(r)$ is already very close to a Gaussian, as shown in Fig.~\ref{fig:volume-j}.
In that figure we plot
\begin{equation}
\mathcal{V}_{\vec{J}}(r)\equiv C f(r),
\label{eq:volume-norm}
\end{equation}
with $C$ a normalization constant such that $\int_0^1 C f(r)\,dr=1$.

The volume $\mathcal{V}_{\vec{J}}$ is defined by
\begin{equation}
\mathcal{V}_{\vec{J}}=\int d\psi \,\delta\!\left(\left\langle \psi\left|\frac{\vec{J}}{j}\right|\psi\right\rangle-\vec{r}\right).
\label{eq:volume-J}
\end{equation}
As noted before, the right-hand side can be interpreted as the probability density of $\vec{r}$.
Thus, to compute the Gaussian approximation, we evaluate the second moment
\begin{align}
\overline{r^2}
& =
\int d\psi \int d^{(3)}r \, r^2 \,
\delta\!\left(\left\langle \psi\left|\frac{\vec{J}}{j}\right|\psi\right\rangle-\vec{r}\right) \\
&=
\int d\psi \left\langle \psi\left|\frac{\vec{J}}{j}\right|\psi\right\rangle^2 .
\end{align}
In Appendix~\ref{app:r2-average} we show that
\begin{equation}
\overline{r^2}=\frac{1}{D-1}.
\label{eq:r2-av}
\end{equation}

The corresponding Gaussian approximation is
\begin{equation}
\mathcal{V}_{\vec{J}}(r)\approx C' \exp\!\left[-\frac{r^2}{2\sigma^2}\right].
\label{eq:gaussian-approx}
\end{equation}
Because \(r^2=x^2+y^2+z^2\) and the distribution is isotropic, the three Cartesian components contribute equally, so that
\begin{equation}
\overline{r^2}=3\sigma^2.
\end{equation}
Using Eq.~\eqref{eq:r2-av}, we then find
\begin{equation}
\sigma^2=\frac{1}{3(D-1)}.
\end{equation}

We have thus computed the volume of a family of pure states constrained by several expectation values. Although the resulting volume depends on a single variable, its piecewise polynomial structure and Gaussian large-dimension limit become apparent only after explicit calculation.

By analogy with the single-observable case studied by Venuti and Zanardi \cite{Venuti2013}, one may expect a multivariate central-limit mechanism for the spin vector \(\vec{r}=(x,y,z)\). If higher-order joint cumulants decay sufficiently rapidly with the Hilbert-space dimension, then \(\mc{V}(x,y,z)\) should become asymptotically Gaussian.This calculation is left for a future publication.

\subsection{Subspace projection}
\label{sec:subs-proy}

\begin{figure}[t]
    \centering
    \includegraphics[width=\columnwidth]{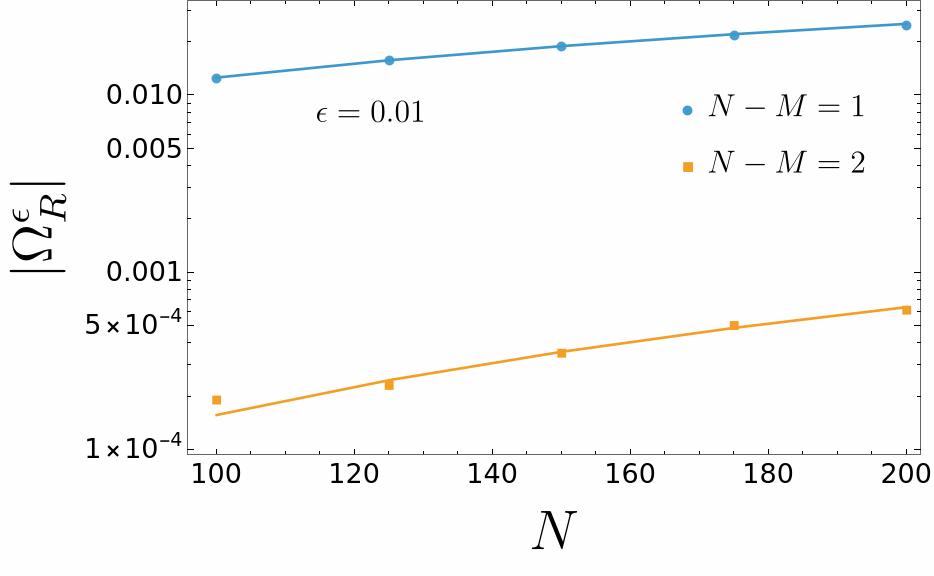}
    \caption{
Regularized volume of the set of states compatible with a subspace of dimension $M$ in an $N$-dimensional Hilbert space, shown as a line. The corresponding numerical results obtained by rejection sampling are shown as dots.
}
    \label{fig:volume-R-eps}
\end{figure}

We start by calculating the volume $\mathcal{V}_R$ associated to a subspace of dimension $M$, 
characterized by a projector $\Pi_R$,
as explained above,
    \begin{equation}
        \mathcal{V}_R = \int d\psi \, \delta(\Pi_R\ket{\psi}-\ket{\psi}).
    \label{eq:volume-R-1}
    \end{equation}
Now, we expand $\ket{\psi}$ in an orthonormal basis, i.e., 
$\ket{\psi}=\sum_{k=1}^N c_k \ket{\phi_k}$, with the first $M$ vectors forming a basis of the 
subspace. 
In terms of the coefficients $c_k$ the Haar measure reads:
    \begin{equation}
        d\psi \propto \prod_{k=1}^N dc_k \, \delta \bigg( \sum_{k=1}^N |c_k|^2-1  \bigg) \, ,
    \label{eq:Haar-ck}
    \end{equation}
where $dc_k=d \Re({c_k}) \, d\Im({c_k})$.
Thus, the integral becomes
    \begin{equation}
       \mathcal{V}_R \propto \int \prod_{k=1}^N dc_k \, \delta \bigg( \sum_{j=1}^N |c_j|^2-1  \bigg)
                                  \prod_{k=M+1}^N \delta^{(2)}(c_k) \, ,
    \label{eq:integral-ck}
    \end{equation}
where we wrote $\propto$ because the measure $\prod_{k=1}^N dc_k $ is not normalized.
Evaluating the integral is an easy task, so we show the final result (including the normalization constant):
    \begin{equation}
        \mathcal{V}_R = \frac{1}{\pi^{N-M}}\frac{(N-1)!}{(M-1)!} \, .
    \label{eq:volume-R}
    \end{equation}
Geometrically, the right-hand side of~\eqref{eq:integral-ck} computes the hypersurface measure of the unit sphere in $2M-1$ dimensions, whereas the normalization constant corresponds to the measure of the unit sphere in $2N-1$ dimensions. The regularized volume, obtained from \eqref{eq:omega-epsilon} with $K=2(N-M)$, is therefore
\begin{equation}
|\Omega_R^\epsilon|
\approx
\left(\frac{4\epsilon^2}{\pi}\right)^{N-M}
\frac{(N-1)!}{(M-1)!}
\label{eq:volume-R-epsilon}
\end{equation}
Figure~\ref{fig:volume-R-eps} shows the regularized volume as function of the total dimension $N$ for two different values of co-dimension ($N-M$), together with the corresponding numerical results obtained by rejection sampling. As the value of the co-dimension increases, i.e., the number of constraints gets larger, the volume rapidly goes to zero, turning a numerical evaluation impractical.

As an alternative to \eqref{eq:integral-ck}, we could have written
    \begin{equation}
       \mathcal{V}^\prime_R \propto \int \prod_{k=1}^N dc_k \, 
       \delta \bigg( \sum_{j=1}^N |c_j|^2-1  \bigg) \,
       \delta \bigg( \sum_{j=1}^M |c_j|^2-1  \bigg) \, ,
    \label{eq:integral-ck-prime-1}
    \end{equation}
because $\delta \bigg( \sum_{j=1}^M |c_j|^2-1  \bigg)$ also defines a subpace of dimension $M$.
The related integral:
    \begin{equation}
       \mathcal{V}^\prime_R(x) \propto \int \prod_{k=1}^N dc_k \, 
       \delta \bigg( \sum_{j=1}^N |c_j|^2-1  \bigg) \,
       \delta \bigg( \sum_{j=1}^M |c_j|^2-x  \bigg) \, ,
    \label{eq:integral-ck-prime}
    \end{equation}
with $0 < x < 1$, coincides with Venuti-Zanardi's $\mc{V}_A(x)$ when $A$ is the projector $\Pi_R$
[see Eq.~(27) in Ref.~\cite{Venuti2013}]. The result is
\begin{equation}
\mathcal{V}^\prime_R(x)
=
\frac{(N-1)!}
{(M-1)!\,(N-M-1)!}
\,x^{M-1}
(1-x)^{N-M-1} \, ,
\label{eq:Vxfactorial}
\end{equation}
thus $x$ is distributed according to a Beta law with parameters $M$ and $N-M$.
Note that
$
\lim_{x\to 1^-}\mathcal{V}^\prime_R(x)=0.
$

This result does not contradict the finite value obtained for $\mathcal{V}_R$. 
The discrepancy originates from the fact that the two quantities are defined with different measures 
on the constrained subspace. 
In particular, the constraint
\begin{equation}
\prod_{k=M+1}^N \delta^{(2)}(c_k)
\end{equation}
localizes the integration directly onto the subspace
$c_{M+1}=\cdots=c_N=0$, 
whereas the constraint
\begin{equation}
\delta\!\left(\sum_{j=M+1}^N |c_j|^2-(1-x)\right)
\end{equation}
induces a radial measure in the orthogonal sector. 
As $x\to 1$, the latter collapses onto the origin with a nontrivial Jacobian factor,
leading to a different limiting behavior. 
Consequently,
$\mathcal{V}_R$ [Eq.~\eqref{eq:volume-R}] cannot be identified with the limit
$\mathcal{V}^\prime_R(x\to 1)$.
See Appendix~\ref{app:equiv} for the general statement of when two constraint
descriptions of a compatible set are equivalent, and for the precise reason
(a critical value of $\langle\psi|\Pi_R|\psi\rangle$) why the two descriptions
compared above are not.

Equation~\eqref{eq:Vxfactorial} also follows from the exact projected
central limit theorem on hyperspheres by Wang, Li and Braunstein~\cite{Braunstein2026}. 
In contrast, $\mathcal{V}_R$
corresponds to the measure of the constraint manifold itself, not to the
distribution of the projected norm.

\subsection{Coarse-graining maps}
\label{sec:coarse-graining-maps}

\subsubsection{Partial trace}

The most iconic example of a coarse-graining map is the partial trace operation, that describes physical systems to which one only has limited access by a reduced state. If we rephrase it in terms of a system in $\mathcal{L}(\mathcal{H}_S)$ of dimension $d_S$, under one’s control, subjected to an environment in $\mathcal{L}(\mathcal{H}_E)$ of dimension $d_E$, the partial trace is a coarse-graining map $\Lambda_{PT}: \, \mathcal{L}(\mathcal{H}_S \otimes \mathcal{H}_E) \rightarrow \mathcal{L}(\mathcal{H}_S)$.

We evaluate the uncertainty associated with an effective description $\rho \in \mc{L}(\mathcal{H}_S)$, obtained by tracing out $\mathcal{H}_E$ in the coarse-graining process. We aim to compute the volume
\begin{equation}
    \mathcal{V}_{\Lambda_{\text{PT}}}(\rho) = \int d\psi \, \delta(\Lambda_{\text{PT}}[\psi] - \rho),
\end{equation}
where $\Lambda_{\text{PT}}[\psi]$ represents the partial trace over the environment $\mathcal{H}_E$ of the state $\psi=\proj{\psi}$. The evaluation of this integral is equivalent to fix $\<\psi| \sigma_i\otimes \idty_E|\psi\>=\tr(\sigma_i \rho)$ for all elements of an operator basis $\{\sigma_i\}$.

We start by making the change of variables $\ket{\psi} \rightarrow U_S \otimes \idty_E \ket{\psi}$, where $U_S$ is an arbitrary unitary acting on $\mathcal{H}_S$. Since the Haar measure is invariant under unitary transformations, it follows that
\begin{align}
    \mathcal{V}_{\Lambda_{\text{PT}}}(\rho) 
       &= \int d \psi \, \delta(U_S \Lambda_{\text{PT}}[\psi]  U_S^{\dagger} - \rho) \\
       &= \int d \psi \, \delta(\Lambda_{\text{PT}}[\psi] - U_S^{\dagger}  \rho U_S).
\end{align}
As $U_S$ is arbitrary, we choose it such that it diagonalizes $\rho$:
\begin{equation}
    \mathcal{V}_{\Lambda_{\text{PT}}}(\rho) = \int d \psi \, \delta(\Lambda_{\text{PT}}[\psi] - \text{diag}[\lambda_1, \dots, \lambda_{d_S}]).
\end{equation}
We assume that all eigenvalues are nonzero, i.e., $\lambda_i \neq 0$ -- the case where some $\lambda_i$ are strictly zero can be obtained by taking the limit $\lambda_i \to 0$.
Now, let us parameterize $\psi$ as
\begin{equation}
    \ket{\psi} = \sum_{i,j} c_{ij} \ket{\phi_i}_S \ket{\gamma_j}_E,
\end{equation}
with $i=1, \dots, d_S$, $j=1, \dots, d_E$, and 
$\braket{\phi_i}{\phi_j} = \braket{\gamma_i}{\gamma_j} = \delta_{ij}$,
such that the reduced state of the system is given by
\begin{equation}
    \rho^S = \tr_E[\ket{\psi}\!\bra{\psi}] = \sum_{ijk} c_{ik} c^*_{jk} \ket{\phi_i}\!\bra{\phi_j}.
\end{equation}
The elements of the reduced density matrix are 
\begin{align}
    \rho^S_{ij} &= \sum_k c_{ik}c^*_{jk}\\
    &= \vec{c}_i \cdot \vec{c}_j^{\, \dagger} \, ,
\end{align}
where we have introduced the vectors of coordinates $\vec{c}_{i(j)}$.
Consequently, the expression for the volume is
\begin{equation}
    \mathcal{V}_{\Lambda_{\text{PT}}}(\rho) \propto \int d \vec{c} \; \prod_{i=1}^{d_S} \delta(\vec{c}_i \vec{c}_i^{\dagger} - \lambda_i) \prod_{i \neq j}^{d_S} \delta(\vec{c}_i \vec{c}_j^{\dagger}),    
\end{equation}
where 
\begin{align}
    d\vec{c} &= \prod_{i=1}^{d_S} d \vec{c}_i, \\
    d \vec{c}_i &= \prod_{k=1}^{d_E} d c_{ik},
\end{align}
and we have omitted a normalization constant depending on the dimensions $d_S$ and $d_E$.
Let us once again make a change of variables $\vec{c}_i = \sqrt{\lambda_i} \vec{z}_i$, such that
\begin{align}
    d\vec{c}_i &= \lambda_i^{d_E} d\vec{z}_i, \\
    d\vec{c} &= (\det[\rho])^{d_E} d\vec{z}, \\
    \prod_{i} \delta(\vec{c}_i \vec{c}_i^{\dagger} - \lambda_i) &= \prod_i \frac{1}{\lambda_i} \delta(\vec{z}_i \vec{z}_i^{\dagger} - 1), \\
    \prod_{i \neq j} \delta(\vec{c}_i \vec{c}_j^{\dagger}) &= \prod_{i \neq j} \frac{1}{\sqrt{\lambda_i \lambda_j}} \delta(\vec{z}_i \vec{z}_j^{\dagger}).
\end{align}
The resulting expression for the volume is
\begin{equation}
\begin{split}
    \mathcal{V}_{\Lambda_{\text{PT}}}(\rho) \propto \det[\rho]^{d_E} \prod_i \frac{1}{\lambda_i} \prod_{i<j} \frac{1}{\lambda_i \lambda_j} \times \\
    \int d \vec{z} \; \prod_{i=1}^{d_S} \delta(\vec{z}_i \vec{z}_i^{\dagger} - 1) \prod_{i \neq j}^{d_S} \delta(\vec{z}_i \vec{z}_j^{\dagger}). 
\end{split}
\end{equation}
Moreover, one can see that
\begin{align}
    \prod_{i} \frac{1}{\lambda_i} &= \frac{1}{\det[\rho]}, \\
    \prod_{i<j} \frac{1}{\lambda_i \lambda_j} &= \frac{1}{(\det[\rho])^{d_S - 1}}.
\end{align}
Finally, we obtain the expression
\begin{align}
\label{eq:vol-pt}
\begin{split}
    \mathcal{V}_{\Lambda_{\text{PT}}}(\rho)  & \propto \det[\rho]^{d_E - d_S} 
    \int d \vec{z} \; \prod_{i=1}^{d_S} \delta(\vec{z}_i \vec{z}_i^{\dagger} - 1) 
                      \prod_{i \neq j}^{d_S} \delta(\vec{z}_i \vec{z}_j^{\dagger}) \\
                      & \propto \det[\rho]^{d_E - d_S}.
\end{split}
\end{align}
The integral above does not depend on $\rho$; it depends only on $d_S$. 
(It represents the volume associated with the identity matrix acting in $\mc{H}_{d_S}$.)

\begin{figure}[H]
  \includegraphics[width=0.5\textwidth,center]{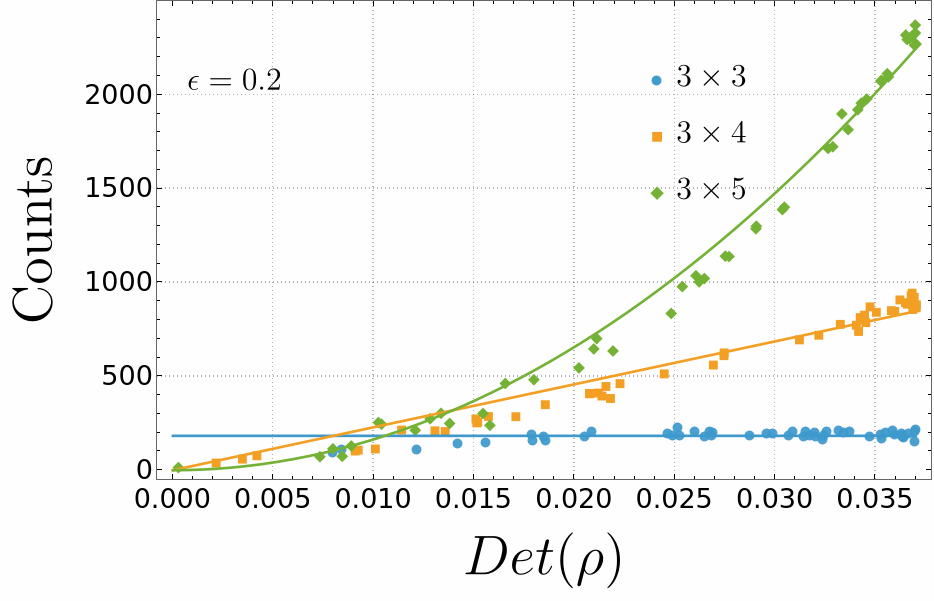}
  \caption{Numerical analysis of the volume of local states. 
  We evaluate how well simulations fit the prediction given by Eq.~\ref{eq:vol-pt}. 
  We considered a set of random local qutrit states. 
  For each $\rho$ in this set, we sampled 100,000 microscopic states $\ket{\psi} \in \mathcal{H}_3 \otimes \mathcal{H}_E$ and checked how many of those satisfied the condition $|\tr(\proj{\psi}\; g_i\otimes \idty_E) - \tr(\rho \; g_i)| < \epsilon$ for all $i\in[8]$, with $g_i$ the usual $i$-th Gell-Mann matrix, and we set $\epsilon = 0.2$. 
  The count of microscopic states obeying these conditions provides an estimate of the volume $\mathcal{V}_{\Lambda_{\text{PT}}}(\rho)$. 
  By plotting the results against $\det(\rho)$, we confirm the expected behavior: for $d_E = 3$ (in blue), the volume of uncertainty is constant, as $d_E = d_S$; for $d_E = 4$, it grows linearly with $\det(\rho)$; and for $d_E = 5$, it grows quadratically.    The indeterminate constant in Eq.~\eqref{eq:vol-pt} was fitted to maximize agreement with the numerical results.}
\label{fig:epsilon}
\end{figure}

As expected, the expression \eqref{eq:vol-pt} accurately reflects the uncertainty in $\rho$. 
One can observe that this volume is maximal for the maximally mixed state, which corresponds, intuitively, to the situation where one knows the least about the possible microscopic state. 
On the other hand, effective states with very high purity --- i.e., one eigenvalue very close to 1 and the remaining eigenvalues very close to 0 --- have a very small corresponding volume. 
In fact, this holds true for any state that is almost rank-deficient, since its determinant would be close to zero.
We point out, however, that such states would never be realistically prepared in the laboratory, since it would require infinite resources \cite{Scharlau18,Masanes17,Schulman05,Wilming17,Clivaz19PRE,Clivaz19PRL}.
Indeed, consider a preparation of state $\rho$ with finite precision $\epsilon$, such that any state $\rho'$ satisfying $|| \rho - \rho'|| < \epsilon$ is equally likely. In particular for rank-deficient $\rho$ (such as a pure state), the only way to ensure that any possible $\rho'$ is also rank-deficient (and thus pure) is by having $\epsilon = 0$.
Moreover, expression~\eqref{eq:vol-pt} predicts that all full-rank states would have the same volume of uncertainty 
if $d_E = d_S$. 
In realistic scenarios, this is unlikely because the environment typically has more degrees of freedom than the system of interest, so $d_E > d_S$ usually holds.
In Fig.~\ref{fig:epsilon}, we provide a numerical analysis further supporting the use of 
$\mathcal{V}_{\Lambda_{\text{PT}}}(\rho)$ as defined above.

Although Ref.~\cite{Ray2023} and our work both adopt a geometric viewpoint, the key difference is the measure: Ray \textit{et al.} define a metric-induced volume on the purification manifold, whereas our partial-trace volume is defined using the Haar measure on the constraint set. Thus, it is not surprising that the resulting expressions differ, most notably in the case $d_S=d_E$, where our volume is independent of $\rho$ while theirs is not.

A further comment: this entire calculation was carried out considering pure microscopic states. It can be easily generalized to mixed states by purifying them using an auxiliary system with dimension $ d_B \geq 1$.
As a result, the corresponding expression for the volume would be
\begin{equation}
\begin{split}
    \mathcal{V}_{\Lambda_{\text{PT}}}^{d_B}(\rho) \propto (\det [\rho])^{d_E d_B-d_S} .
\end{split}
\end{equation}

\subsubsection{Detector coarse-graining map}

Another example of a coarse-graining map, of particular interest to us, is the one describing the measurements performed on an imperfect detector. Consider the usual setup of a lattice of cold atoms, trapped in an optical cavity \cite{lewenstein2012ultracold}. Consider the individual atoms as 3-level systems, which can be in states $\ket{0}$, $\ket{1}$ or $\ket{2}$. In a very simplified model, the apparatus that performs the corresponding measurements may not be able to resolve, for instance, the levels $\ket{1}$ and $\ket{2}$, being only able to ascertain whether the atom is in the ground state or not. This inherent imprecision in the measurements will result in a coarse-grained description of the system in terms of only two effective levels: $\ket{0}$, to which atoms in the
ground state are assigned, and $\ket{1}$, to which the two remaining states are assigned. To characterize the map $\Lambda_D$ describing the action of this imperfect device, as in \cite{cris2017,pedrinho}, we recall that the action of a completely positive map $\Phi$ on any quantum state $\rho$ can be characterized in terms of operators $\{K_i\}_i$ satisfying $\sum_i = K^{\dagger}_i K_i = \idty$ such that

\begin{equation}
    \Phi[\rho] = \sum_i K_i \rho K_i^{\dagger}.
\end{equation}

Thus, the Kraus operators that describe $\Lambda_D$ are

\begin{equation}\label{kraus-ld}
    K_1 = \begin{bmatrix}
1 & 0 & 0\\
0 & 1/\sqrt{2}& 1/\sqrt{2}
\end{bmatrix}, \, K_2 = \begin{bmatrix}
0 & 0 & 0\\
0 & 1/\sqrt{2}& -1/\sqrt{2}
\end{bmatrix}.
\end{equation}

The action of this coarse-graining map is, alternatively, given by

\begin{center}
\begin{tabular}{l}
$\Lambda_D[\ket{0}\!\bra{0}]=\ket{0}\!\bra{0}$, \\
$\Lambda_D[\ket{0}\!\bra{1}]=\frac{1}{\sqrt{2}}\ket{0}\!\bra{1}$, \\
$\Lambda_D[\ket{0}\!\bra{2}]=\frac{1}{\sqrt{2}}\ket{0}\!\bra{1}$, \\
$\Lambda_D[\ket{1}\!\bra{0}]=\frac{1}{\sqrt{2}}\ket{1}\!\bra{0}$, \\
$\Lambda_D[\ket{1}\!\bra{1}]=\ket{1}\!\bra{1}$, \\
$\Lambda_D[\ket{1}\!\bra{2}]=0$, \\
$\Lambda_D[\ket{2}\!\bra{0}]=\frac{1}{\sqrt{2}}\ket{1}\!\bra{0}$, \\
$\Lambda_D[\ket{2}\!\bra{1}]=0$, \\
$\Lambda_D[\ket{2}\!\bra{2}]=\ket{1}\!\bra{1}$.
\end{tabular}
\end{center}

Notice that the map reflects the fact that coherence terms within the excited subspace (that is, the space spanned by $\{\ket{1}, \ket{2}\}$) must vanish, since they cannot be discriminated after the coarse-grained mapped has been applied. 

Once the calculation for the case of the partial trace have been done, 
one can similarly evaluate the uncertainty associated to an effective description $\rho$ when the coarse-graining map is $\Lambda_D$, which yields
\begin{equation}\label{vol-d}
\begin{split}
    \mathcal{V}^{d_B}_{\Lambda_D}(\rho) = 2^{4 - 3 d_B} \pi^{1+4d_B}(1 + z)^{-d_B}  \times \\
    \frac{(1 - x^2 - y^2 - z^2)^{2 (d_B-1)} }{\Gamma[d_B] \Gamma[2 d_B - 1]},
\end{split}
\end{equation}
where $x,y,z$ are the Bloch vectors of $\rho$. 
Moreover, here we also consider mixed states, which are purified with an auxiliary system of dimension $d_B$. 
This calculation is significantly more complicated, and the details are presented in Appendix~\ref{app:vol-Lambda}, 
where we have adapted a calculation lead in \cite{vallejos2022}.

	 
\section{Concluding remarks}
\label{sec:conclusions}

In this work, we introduced a quantum counterpart of Boltzmann’s counting procedure by defining an entropic quantity from the Hilbert-space volume of pure states compatible with a given preparation. This construction restores, in the quantum setting, the twofold role played by a macroregion in classical statistical mechanics. Combined with the average assignment map introduced in our previous works, the same set of compatible microscopic states can now be used both to assign a statistical state to a preparation and to quantify the remaining microscopic uncertainty through its volume. We evaluated this volume for several paradigmatic preparation constraints, including restrictions to a subspace, fixed expectation values, and subsystem descriptions induced by coarse-graining maps, obtaining explicit expressions or scaling laws in representative cases. Since the construction is tied to the preparation rather than to equilibrium, it also applies naturally to nonequilibrium situations.

Throughout this work, each preparation is represented by a fixed choice of constraint functions. As discussed in Appendix~\ref{app:equiv}, alternative but mathematically equivalent definitions of the same compatible set modify the corresponding compatible-set volume only by an overall constant Jacobian factor, and hence shift the associated entropy by an additive constant. This clarifies the mathematical status of the compatible-set volume employed throughout the present work.

Several questions remain open. On the mathematical side, it will be important to establish the general properties of the proposed entropy, including its additivity, subadditivity, continuity, and behavior under composition and coarse graining. 
It would also be interesting to investigate whether one can formulate an intrinsic notion of compatible-set volume independent of the particular defining equations of the constraint manifold. On the physical side, the framework can be used to investigate thermodynamic processes by following how the set of compatible microscopic states—and hence its entropy—changes under dynamics, measurements, and finite-resolution interventions. In particular, identifying general conditions under which this volume increases may provide a route toward a microscopic and mathematically controlled formulation of entropy production and the second law of thermodynamics.

\section{Acknowledgements} 
This work is supported in part by the National Council for Scientific and Technological Development, 
CNPq Brazil (projects: Universal Grants No. 406499/2021-7 and 408990/2025-2, and projects 305071/2022-0 
and 409611/2022-0), 
the Carlos Chagas Foundation for Research Support of the State of Rio de Janeiro (FAPERJ, Grant APQ1 E-26/210.576/2024),
and it is part of the National Institute of Science and Technology for Applied Quantum Computing through CNPq process No. 408884/2024-0. I.V. acknowledges funding from INRIA and CIEDS through the Action Exploratoire project DEPARTURE. Part of this work was carried out while F.B. was a faculty member at the Instituto de Física de São Carlos, Universidade de São Paulo, São Carlos, São Paulo, Brazil.

\appendix

\section{Average of \texorpdfstring{$r^2$}{r²} for the spin-\texorpdfstring{$j$}{j} Bloch vector}
\label{app:r2-average}

In this appendix we compute the average value of $r^2$ associated with the vector
\begin{equation}
\vec r = \left(r_1,r_2,r_3\right), \qquad r_i=\left\langle \psi \left| \frac{J_i}{j} \right| \psi \right\rangle ,
\end{equation}
where $\vec J=(J_1,J_2,J_3)$ is the angular-momentum operator in the irreducible spin-$j$ representation, with block dimension $D=2j+1$. The average is taken with respect to the unitarily invariant measure on pure states in the corresponding Hilbert space.

The quantity of interest is
\begin{equation}
\overline{r^2}
=
\int d\psi\, r^2
=
\int d\psi \sum_{i=1}^3 \left\langle \psi \left| \frac{J_i}{j} \right| \psi \right\rangle^2 .
\end{equation}
Using the standard Haar-average identity for two expectation values of operators $A$ and $B$ in a $D$-dimensional Hilbert space \cite{Mele2024Haar},
\begin{equation}
\int d\psi \,\langle \psi|A|\psi\rangle \langle \psi|B|\psi\rangle
=
\frac{\mathrm{tr}(AB)+\mathrm{tr}(A)\mathrm{tr}(B)}{D(D+1)},
\end{equation}
and noting that $\mathrm{tr}(J_i)=0$, we obtain
\begin{equation}
\overline{r^2}
=
\frac{1}{j^2}\sum_{i=1}^3 \frac{\mathrm{tr}(J_i^2)}{D(D+1)}.
\end{equation}
By rotational symmetry in the irreducible representation,
\begin{equation}
\mathrm{tr}(J_1^2)=\mathrm{tr}(J_2^2)=\mathrm{tr}(J_3^2),
\end{equation}
and since
\begin{equation}
J_1^2+J_2^2+J_3^2=j(j+1)\,\mathbb{I},
\end{equation}
taking the trace yields
\begin{equation}
\sum_{i=1}^3 \mathrm{tr}(J_i^2)=j(j+1)D.
\end{equation}
Therefore,
\begin{equation}
\overline{r^2}
=
\frac{1}{j^2}\frac{j(j+1)D}{D(D+1)}
=
\frac{j+1}{j(D+1)}.
\end{equation}
Finally, using $D=2j+1$, we obtain
\begin{equation}
\overline{r^2}=\frac{1}{D-1}.
\end{equation}
This is the desired result.


\onecolumngrid

\section{Calculation of the volume associated with the detector channel}
\label{app:vol-Lambda}

To obtain Eq.~\eqref{vol-d}, we will follow calculations completely analogous to those found in \cite{vallejos2022}, letting Mathematica do the brute-force evaluations \cite{Mathematica}.
Let us start by decomposing a given effective state $\rho \in \mathcal{L}(\mathcal{H}_2)$ as
\begin{equation}\label{eq:1}
    \rho = \begin{pmatrix}
 \rho_{00} & 0\\
 0 & \rho_{11}
 \end{pmatrix} + \frac{1}{2}(x \sigma_x + y \sigma_y),
\end{equation}
where
$x=\tr[\rho \sigma_x]$, $y=\tr[\rho \sigma_y]$, and $\rho_{ij}$
with
$i,j \in \{0,1\}$ are the matrix coefficients of $\rho$.
Let us parameterize the matrix elements of microscopic pure states $\psi \in \mathcal{L}(\mathcal{H}_3)$ as
$\psi_{ij} = c_i c_j^*$, $i,j \in \{1,2,3\}$, such that $\psi = c c^{\dagger}$.
We can write the action of $\Lambda_D$ on a generic state as

\begin{equation}\label{eq:2}
    \Lambda_D[\psi] = \begin{pmatrix}
    c_{1}c_1^* & c_{1}(c_2^*+c_3^*)/\sqrt{2}\\
    c_{1}^*(c_2+c_3)/\sqrt{2} & c_{2}c_2^* + c_{3}c_3^*
                      \end{pmatrix} ,
\end{equation}
where we have simply applied the Kraus operators defined in \eqref{kraus-ld}.
In terms of these parameters, we can write the volume of $\Omega_{\Lambda_D}(\rho)$ as
\begin{equation}
    \mathcal{V}_{\Lambda_D}(\rho) = \int d\mu_{\psi}
             \delta(c_1 c_1^* - \rho_{00})
             \delta(c_2 c_2^* + c_3 c_3^* - \rho_{11})
             \delta(\tr[\Lambda_D[\psi]\sigma_x] - x)
             \delta(\tr[\Lambda_D[\psi]\sigma_y] - y),
\end{equation}
where we have simply used \Cref{eq:1,eq:2} to impose the constraint $\Lambda_D[\psi] = \rho$.

The trick now is to consider the Laplace or Fourier transforms of these objects,
perform some calculations on the results, and subsequently evaluate the corresponding inverse transforms.
The advantage of this method is that, between applying the integral transform and its inverse,
the expression will be independent of $\rho$, which will greatly simplify the calculations.

We can rewrite each term in parentheses as follows:
\begin{equation}
    \begin{split}
        \delta(c_1 c_1^* - \rho_{00}) &= \mathfrak{L}^{-1}\{\mathfrak{L} \{\delta(c_1 c_1^* - \rho_{00})\}(s_0) \}(\rho_{00}) \\
        &= \mathfrak{L}^{-1}\{ \int_0^{\infty} d\rho_{00} \, e^{-s_0 \rho_{00}}\delta(c_1 c_1^* - \rho_{00}) \}(\rho_{00}) \\
        &= \mathfrak{L}^{-1}\{ e^{-s_0 c_1 c_1^*} \}(\rho_{00}),
    \end{split}
\end{equation}
and similarly
\begin{equation}
    \begin{split}
        \delta(c_2 c_2^* + c_3 c_3^* - \rho_{11}) = \mathfrak{L}^{-1}\{ e^{-s_1 (c_2 c_2^* + c_3 c_3^*)} \}(\rho_{11}),
    \end{split}
\end{equation}
where $\mathfrak{L}$ is the Laplace transform. Furthermore,
\begin{equation}
    \begin{split}
        \delta(\tr[\Lambda_D[\psi]\sigma_x] - x) &= \mathfrak{F}^{-1}\{\mathfrak{F} \{\delta(\tr[\Lambda_D[\psi]\sigma_x] - x)\}(k_x) \}(x) \\
        &= \mathfrak{F}^{-1}\{ \int_{-\infty}^{+\infty} dx \, e^{-i k_x x}\delta(\tr[\Lambda_D[\psi]\sigma_x] - x) \}(x) \\
        &= \mathfrak{F}^{-1}\{ e^{-i k_x \tr[\Lambda_D[\psi]\sigma_x]} \}(x) \\
        &= \mathfrak{F}^{-1}\{ e^{-i k_x c\Lambda'_D[\sigma_x]c^{\dagger}} \}(x),
    \end{split}
\end{equation}
where $\Lambda'_D$ is the dual of the map $\Lambda_D$, and similarly
\begin{equation}
    \begin{split}
        \delta(\tr[\Lambda_D[\psi]\sigma_y] - y) = \mathfrak{F}^{-1}\{ e^{-i k_y c\Lambda'_D[\sigma_y]c^{\dagger}} \}(y),
    \end{split}
\end{equation}
where $\mathfrak{F}$ is the Fourier transform
(which we perform here instead of the Laplace transform because $x,y$ can assume negative values).
Now, denote this intermediate object obtained after applying the integral transform by $\mathcal{I}$,
and rewrite it as
\begin{equation}
    \mathcal{I}[s_0,s_1,k_x,k_y] = \int d(c) \, e^{-c^{\dagger}A c},
\end{equation}
with
\begin{equation}
    A = \mathrm{diag}(s_0,s_1,s_1) + i \Lambda_D'[k_x \sigma_x + k_y \sigma_y]
      = \begin{pmatrix}
            s_0                         & \frac{ik_x + k_y}{\sqrt{2}} & \frac{ik_x + k_y}{\sqrt{2}}\\
            \frac{ik_x - k_y}{\sqrt{2}} & s_1                         & 0 \\
            \frac{ik_x - k_y}{\sqrt{2}} & 0                           & s_1
        \end{pmatrix}.
\end{equation}
The Gaussian integral $\mathcal{I}$ can be readily evaluated  \cite{Altland-Simons2010}:
\begin{equation}
    \mathcal{I}[s_0,s_1,k_x,k_y] = \frac{\pi^4}{\det A}.
\end{equation}
One can now consider $\psi$ to be mixed, and let it be purified by an auxiliary space of dimension $d_B$.
Then, a similar calculation yields
\begin{equation}
    \mathcal{I}_{d_B}[s_0,s_1,k_x,k_y] = \frac{\pi^{4 d_B}}{(\det A)^{d_B}}.
\end{equation}
All that is left is to perform the inverse transforms, so that we recover the volume of uncertainty of $\rho$.
Once again, with the aid of Mathematica \cite{Mathematica}, we obtain the final expression in terms of the Bloch vector of $\rho$:
\begin{equation}
    \mathcal{V}_{\Lambda_D}(\rho) =
       \frac{ 2^{4 - 3 d_B} \pi^{1+4d_B} }{ \Gamma[d_B] \Gamma[2 d_B - 1] }
       \frac{(1 - x^2 - y^2 - z^2)^{2 (d_B-1)}}{(1 + z)^{d_B}} .
\end{equation}

\twocolumngrid

\section{Equivalent definitions of the compatible set}
\label{app:equiv}

For definiteness, we consider a compatible set defined by two constraint
functions,
\begin{equation}
\Omega
=
\{\psi\in\mathcal{H}:f_1(\psi)=0,\;f_2(\psi)=0\},
\label{eq:A1}
\end{equation}
although the discussion applies equally to an arbitrary number of
constraints, including the case of a single constraint. The functions
$f_1$ and $f_2$ represent the physical constraints imposed by the
preparation. Depending on the preparation under consideration, they may
encode, for example, expectation-value restrictions, subspace constraints,
or other conditions defining the set of compatible pure states.

The compatible-set volume is defined by

\begin{equation}
V
=
\int d\psi\,
\delta(f_1(\psi))
\delta(f_2(\psi)),
\label{eq:A4}
\end{equation}
where $d\psi$ denotes the Haar measure over pure states.

The same subset $\Omega$ may equivalently be described by any pair of smooth
functions $F,G:\mathbb R^2\to\mathbb R$ such that

\begin{equation}
F(f_1(\psi),f_2(\psi))=0,
\qquad
G(f_1(\psi),f_2(\psi))=0,
\label{eq:A5}
\end{equation}
where $F,G$ define the same set $\Omega$ as $f_1,f_2$ provided

\begin{equation}
F(0,0)=G(0,0)=0,
\label{eq:A6}
\end{equation}
and $(0,0)$ is the only common zero of $F,G$ over the range of values
$(f_1,f_2)$ actually attains. This holds automatically for any bona fide
change of variables -- rescalings, translations, or, more generally, genuine
diffeomorphisms of the constraint values. We further require the zero to be
\emph{transversal},

\begin{equation}
\det
\left(
\frac{\partial(F,G)}
{\partial(f_1,f_2)}
\right)_{(0,0)}
\neq0,
\label{eq:A7}
\end{equation}
which will be needed below for the volume transformation law.

The corresponding compatible-set volume defined by $F,G$ is

\begin{equation}
V'
=
\int
d\psi\,
\delta(F)
\delta(G).
\label{eq:A8}
\end{equation}
The multidimensional transformation law for Dirac delta distributions gives

\begin{equation}
\delta(F)\delta(G)
=
\frac{
\delta(f_1)\delta(f_2)
}{
\left|
\det
\left(
\frac{\partial(F,G)}
{\partial(f_1,f_2)}
\right)_{(0,0)}
\right|
},
\label{eq:A9}
\end{equation}
where the Jacobian is evaluated at $(f_1,f_2)=(0,0)$. Consequently,

\begin{equation}
V'
=
\frac{
V
}{
\left|
\det
\left(
\frac{\partial(F,G)}
{\partial(f_1,f_2)}
\right)_{(0,0)}
\right|
}.
\label{eq:A10}
\end{equation}

As an illustration of why Eq.~\eqref{eq:A10} does not apply indiscriminately,
consider the vector restriction $c_{M+1}=\cdots=c_N=0$ used to define
$\Omega_R$ in Sec.~\ref{sec:subs-proy} -- a set of $2(N-M)$ real equations --
compared with the single average-value constraint
$\langle\psi|\Pi_R|\psi\rangle=1$ of Ref.~\cite{Venuti2013}. Since the two
descriptions involve a different numbers of constraints,
$\partial(F,G)/\partial(f_1,f_2)$ is not even a well-defined square matrix in
this case, so condition \eqref{eq:A7} does not directly apply. The deeper
reason the two volumes remain inequivalent even in a matched single-constraint
comparison is that $\langle\psi|\Pi_R|\psi\rangle=1$ is an instance of the
fixed expectation-value class of Eq.~\eqref{eq:Omega-a-00} with $A=\Pi_R$,
evaluated at $a=1$, the \emph{maximal eigenvalue} of $\Pi_R$ and hence a
critical value of the quadratic form $\langle\psi|A|\psi\rangle$. At such a
critical value the level set fails to be a regular codimension-one
hypersurface and instead collapses onto $\Omega_R$ itself, which is why
$\mathcal V_R$ [Eq.~\eqref{eq:volume-R}] and $\lim_{x\to1^-}\mathcal
V_R'(x)$ [Eq.~\eqref{eq:Vxfactorial}] need not, and do not, coincide.

\bibliographystyle{unsrtnat}

\bibliography{ref} 

\begin{thebibliography}{33}
\providecommand{\natexlab}[1]{#1}
\providecommand{\url}[1]{\texttt{#1}}
\expandafter\ifx\csname urlstyle\endcsname\relax
  \providecommand{\doi}[1]{doi: #1}\else
  \providecommand{\doi}{doi: \begingroup \urlstyle{rm}\Url}\fi

\bibitem[Correia et~al.(2021)Correia, Obando, Vallejos, and de~Melo]{correia2021}
Pedro~Silva Correia, Paola~Concha Obando, Ra\'ul~O. Vallejos, and Fernando de~Melo.
\newblock Macro-to-micro quantum mapping and the emergence of nonlinearity.
\newblock \emph{Phys. Rev. A}, 103:\penalty0 052210, May 2021.
\newblock \doi{10.1103/PhysRevA.103.052210}.
\newblock URL \url{https://link.aps.org/doi/10.1103/PhysRevA.103.052210}.

\bibitem[Vallejos et~al.(2022)Vallejos, Correia, Obando, O'Neill, Tacla, and de~Melo]{vallejos2022}
Ra{\'u}l~O Vallejos, Pedro~Silva Correia, Paola~Concha Obando, Nina~Machado O'Neill, Alexandre~B Tacla, and Fernando de~Melo.
\newblock Quantum state inference from coarse-grained descriptions: Analysis and an application to quantum thermodynamics.
\newblock \emph{Physical Review A}, 106\penalty0 (1):\penalty0 012219, 2022.

\bibitem[Ray et~al.(2023)Ray, Alsing, Cafaro, and Jacinto]{Ray2023}
Shannon Ray, Paul~M. Alsing, Carlo Cafaro, and H~S. Jacinto.
\newblock A differential-geometric approach to quantum ignorance consistent with entropic properties of statistical mechanics.
\newblock \emph{Entropy}, 25\penalty0 (5):\penalty0 788, May 2023.
\newblock ISSN 1099-4300.
\newblock \doi{10.3390/e25050788}.
\newblock URL \url{http://dx.doi.org/10.3390/e25050788}.

\bibitem[{\v{S}}afr{\'a}nek et~al.(2020){\v{S}}afr{\'a}nek, Aguirre, and Deutsch]{Safranek2020}
Dominik {\v{S}}afr{\'a}nek, Anthony Aguirre, and J.~M. Deutsch.
\newblock Classical dynamical coarse-grained entropy and comparison with the quantum version.
\newblock \emph{Physical Review E}, 102:\penalty0 032106, 2020.
\newblock \doi{10.1103/PhysRevE.102.032106}.

\bibitem[{\v{S}}afr{\'a}nek et~al.(2021){\v{S}}afr{\'a}nek, Aguirre, Schindler, and Deutsch]{Safranek2021}
Dominik {\v{S}}afr{\'a}nek, Anthony Aguirre, Joseph Schindler, and J.~M. Deutsch.
\newblock A brief introduction to observational entropy.
\newblock \emph{Foundations of Physics}, 51:\penalty0 101, 2021.
\newblock \doi{10.1007/s10701-021-00498-x}.

\bibitem[{\v{S}}afr{\'a}nek et~al.(2019{\natexlab{a}}){\v{S}}afr{\'a}nek, Deutsch, and Aguirre]{Safranek2019a}
David {\v{S}}afr{\'a}nek, James~M. Deutsch, and Anthony Aguirre.
\newblock Quantum coarse-grained entropy and thermodynamics.
\newblock \emph{Physical Review A}, 99:\penalty0 010101, 2019{\natexlab{a}}.
\newblock \doi{10.1103/PhysRevA.99.010101}.

\bibitem[{\v{S}}afr{\'a}nek et~al.(2019{\natexlab{b}}){\v{S}}afr{\'a}nek, Deutsch, and Aguirre]{Safranek2019b}
David {\v{S}}afr{\'a}nek, James~M. Deutsch, and Anthony Aguirre.
\newblock Quantum coarse-grained entropy and thermalization in closed systems.
\newblock \emph{Physical Review A}, 99:\penalty0 012103, 2019{\natexlab{b}}.
\newblock \doi{10.1103/PhysRevA.99.012103}.

\bibitem[Popescu et~al.(2006)Popescu, Short, and Winter]{Popescu2006}
Sandu Popescu, Anthony~J. Short, and Andreas Winter.
\newblock Entanglement and the foundations of statistical mechanics.
\newblock \emph{Nature Physics}, 2\penalty0 (11):\penalty0 754--758, 2006.
\newblock \doi{10.1038/nphys444}.

\bibitem[M{\"u}ller et~al.(2011)M{\"u}ller, Gross, and Eisert]{MullerGrossEisert2011}
Markus~P. M{\"u}ller, David Gross, and Jens Eisert.
\newblock Concentration of measure for quantum states with a fixed expectation value.
\newblock \emph{New Journal of Physics}, 13\penalty0 (7):\penalty0 073025, 2011.
\newblock \doi{10.1088/1367-2630/13/7/073025}.

\bibitem[Bartsch and Gemmer(2009)]{BartschGemmer2009}
Christian Bartsch and Jochen Gemmer.
\newblock Dynamical typicality of quantum expectation values.
\newblock \emph{Physical Review Letters}, 102\penalty0 (11):\penalty0 110403, 2009.
\newblock \doi{10.1103/PhysRevLett.102.110403}.

\bibitem[Reimann(2018)]{Reimann2018}
Peter Reimann.
\newblock Dynamical typicality of isolated many-body quantum systems.
\newblock \emph{Physical Review E}, 97\penalty0 (6):\penalty0 062129, 2018.
\newblock \doi{10.1103/PhysRevE.97.062129}.

\bibitem[Brody and Hughston(1998)]{Brody1998}
D.~C. Brody and L.~P. Hughston.
\newblock The quantum canonical ensemble.
\newblock \emph{Journal of Mathematical Physics}, 39:\penalty0 6502, 1998.

\bibitem[Bender et~al.(2005)Bender, Brody, and Hook]{Bender2005}
C.~M. Bender, D.~C. Brody, and D.~W. Hook.
\newblock Solvable model of quantum microcanonical states.
\newblock \emph{Journal of Physics A: Mathematical and General}, 38:\penalty0 L607, 2005.

\bibitem[Brody et~al.(2007{\natexlab{a}})Brody, Hook, and Hughston]{Brody2007}
D.~C. Brody, D.~W. Hook, and L.~P. Hughston.
\newblock Quantum phase transitions without thermodynamic limits.
\newblock \emph{Proceedings of the Royal Society A: Mathematical, Physical and Engineering Science}, 463:\penalty0 2021, 2007{\natexlab{a}}.

\bibitem[Brody et~al.(2007{\natexlab{b}})Brody, Hook, and Hughston]{Brody2007b}
Dorje~C Brody, Daniel~W Hook, and Lane~P Hughston.
\newblock On quantum microcanonical equilibrium.
\newblock \emph{Journal of Physics: Conference Series}, 67:\penalty0 012025, May 2007{\natexlab{b}}.
\newblock ISSN 1742-6596.
\newblock \doi{10.1088/1742-6596/67/1/012025}.
\newblock URL \url{http://dx.doi.org/10.1088/1742-6596/67/1/012025}.

\bibitem[Dunkl et~al.(2011{\natexlab{a}})Dunkl, Gawron, Holbrook, Pucha{\l}a, and {\.Z}yczkowski]{Dunkl2011}
C.~F. Dunkl, P.~Gawron, J.~A. Holbrook, Z.~Pucha{\l}a, and K.~{\.Z}yczkowski.
\newblock Numerical shadows: Measures and densities on the numerical range.
\newblock \emph{Linear Algebra and its Applications}, 434:\penalty0 2042, 2011{\natexlab{a}}.

\bibitem[Dunkl et~al.(2011{\natexlab{b}})Dunkl, Gawron, Holbrook, Miszczak, Pucha{\l}a, and {\.Z}yczkowski]{Dunkl2011b}
C.~F. Dunkl, P.~Gawron, J.~A. Holbrook, J.~A. Miszczak, Z.~Pucha{\l}a, and K.~{\.Z}yczkowski.
\newblock Numerical shadow and geometry of quantum states.
\newblock \emph{Journal of Physics A: Mathematical and Theoretical}, 44:\penalty0 335301, 2011{\natexlab{b}}.

\bibitem[Pucha{\l}a et~al.(2012)Pucha{\l}a, Miszczak, Gawron, Dunkl, Holbrook, and {\.Z}yczkowski]{Puchal2012}
Z.~Pucha{\l}a, J.~A. Miszczak, P.~Gawron, C.~F. Dunkl, J.~A. Holbrook, and K.~{\.Z}yczkowski.
\newblock Restricted numerical shadow and the geometry of quantum entanglement.
\newblock \emph{Journal of Physics A: Mathematical and Theoretical}, 45:\penalty0 415309, 2012.

\bibitem[Venuti and Zanardi(2013)]{Venuti2013}
Lorenzo~Campos Venuti and Paolo Zanardi.
\newblock Probability density of quantum expectation values.
\newblock \emph{Physics Letters A}, 377\penalty0 (31–33):\penalty0 1854--1861, 2013.
\newblock \doi{10.1016/j.physleta.2013.05.041}.
\newblock URL \url{https://doi.org/10.1016/j.physleta.2013.05.041}.

\bibitem[Saideh et~al.(2015)Saideh, Ribeiro, Ferrini, Coudreau, Milman, and Keller]{ibrahim}
Ibrahim Saideh, A.~D. Ribeiro, Giulia Ferrini, Thomas Coudreau, P\'erola Milman, and Arne Keller.
\newblock General dichotomization procedure to provide qudit entanglement criteria.
\newblock \emph{Phys. Rev. A}, 92:\penalty0 052334, Nov 2015.
\newblock \doi{10.1103/PhysRevA.92.052334}.
\newblock URL \url{https://link.aps.org/doi/10.1103/PhysRevA.92.052334}.

\bibitem[Wang et~al.(2026)Wang, Li, and Braunstein]{Braunstein2026}
Zhi-Wei Wang, Pei-Wen Li, and Samuel~L. Braunstein.
\newblock Exact geometric typicality and bipartite entanglement from the projected central limit theorem on hyperspheres.
\newblock \emph{arXiv preprint}, 2026.
\newblock URL \url{https://arxiv.org/abs/2605.29732}.

\bibitem[Scharlau and M{\"u}ller(2018)]{Scharlau18}
Jakob Scharlau and Markus~P. M{\"u}ller.
\newblock Quantum horn's lemma, finite heat baths, and the third law of thermodynamics.
\newblock \emph{Quantum}, 2:\penalty0 54, 2018.
\newblock \doi{10.22331/q-2018-10-04-54}.
\newblock URL \url{https://quantum-journal.org/papers/q-2018-10-04-54/}.

\bibitem[Masanes and Oppenheim(2017)]{Masanes17}
Lluis Masanes and Jonathan Oppenheim.
\newblock A general derivation and quantification of the third law of thermodynamics.
\newblock \emph{Nature Communications}, 8:\penalty0 14538, 2017.
\newblock \doi{10.1038/ncomms14538}.
\newblock URL \url{https://www.nature.com/articles/ncomms14538}.

\bibitem[Schulman et~al.(2005)Schulman, Mor, and Weinstein]{Schulman05}
Leonard~J. Schulman, Tal Mor, and Yossi Weinstein.
\newblock Physical limits of heat-bath algorithmic cooling.
\newblock \emph{Physical Review Letters}, 94:\penalty0 120501, 2005.
\newblock \doi{10.1103/PhysRevLett.94.120501}.

\bibitem[Wilming and Gallego(2017)]{Wilming17}
Henrik Wilming and Rodrigo Gallego.
\newblock Third law of thermodynamics as a single inequality.
\newblock \emph{Physical Review X}, 7:\penalty0 041033, 2017.
\newblock \doi{10.1103/PhysRevX.7.041033}.

\bibitem[Clivaz et~al.(2019{\natexlab{a}})Clivaz, Silva, Haack, Brask, Brunner, and Huber]{Clivaz19PRE}
Fabien Clivaz, Ralph Silva, G{\'e}raldine Haack, Jonatan~Bohr Brask, Nicolas Brunner, and Marcus Huber.
\newblock Unifying paradigms of quantum refrigeration: fundamental limits of cooling and associated work costs.
\newblock \emph{Physical Review E}, 100:\penalty0 042130, 2019{\natexlab{a}}.
\newblock \doi{10.1103/PhysRevE.100.042130}.

\bibitem[Clivaz et~al.(2019{\natexlab{b}})Clivaz, Silva, Haack, Brask, Brunner, and Huber]{Clivaz19PRL}
Fabien Clivaz, Ralph Silva, G{\'e}raldine Haack, Jonatan~Bohr Brask, Nicolas Brunner, and Marcus Huber.
\newblock Unifying paradigms of quantum refrigeration: A universal and attainable bound on cooling.
\newblock \emph{Physical Review Letters}, 123:\penalty0 170605, 2019{\natexlab{b}}.
\newblock \doi{10.1103/PhysRevLett.123.170605}.

\bibitem[Lewenstein et~al.(2012)Lewenstein, Sanpera, and Ahufinger]{lewenstein2012ultracold}
Maciej Lewenstein, Anna Sanpera, and Veronica Ahufinger.
\newblock \emph{Ultracold Atoms in Optical Lattices: Simulating quantum many-body systems}.
\newblock Oxford University Press (UK), 2012.

\bibitem[Duarte et~al.(2017)Duarte, Carvalho, Bernardes, and de~Melo]{cris2017}
Cristhiano Duarte, Gabriel~Dias Carvalho, Nadja~K. Bernardes, and Fernando de~Melo.
\newblock Emerging dynamics arising from coarse-grained quantum systems.
\newblock \emph{Phys. Rev. A}, 96:\penalty0 032113, Sep 2017.
\newblock \doi{10.1103/PhysRevA.96.032113}.
\newblock URL \url{https://link.aps.org/doi/10.1103/PhysRevA.96.032113}.

\bibitem[Silva~Correia and de~Melo(2019)]{pedrinho}
Pedro Silva~Correia and Fernando de~Melo.
\newblock Spin-entanglement wave in a coarse-grained optical lattice.
\newblock \emph{Phys. Rev. A}, 100:\penalty0 022334, Aug 2019.
\newblock \doi{10.1103/PhysRevA.100.022334}.
\newblock URL \url{https://link.aps.org/doi/10.1103/PhysRevA.100.022334}.

\bibitem[Mele(2024)]{Mele2024Haar}
Antonio~Anna Mele.
\newblock Introduction to haar measure tools in quantum information: A beginner's tutorial.
\newblock \emph{Quantum}, 8:\penalty0 1340, 2024.
\newblock \doi{10.22331/q-2024-05-08-1340}.
\newblock URL \url{https://quantum-journal.org/papers/q-2024-05-08-1340/}.

\bibitem[{Wolfram Research, Inc.}(2024)]{Mathematica}
{Wolfram Research, Inc.}
\newblock Mathematica.
\newblock Computer software, 2024.
\newblock URL \url{https://www.wolfram.com/mathematica/}.

\bibitem[Altland and Simons(2010)]{Altland-Simons2010}
Alexander Altland and Ben~D. Simons.
\newblock \emph{Condensed Matter Field Theory}.
\newblock Cambridge University Press, Cambridge, 2 edition, 2010.
\newblock ISBN 9780521769754.

\end{thebibliography}
	
\end{document}